\documentclass[aps,floats,subeqn,showpacs,amssymb]{revtex4}
\usepackage{epsfig}

\newcommand{\be}{\begin{equation}}
\newcommand{\ee}{\end{equation}}
\newcommand{\bea}{\begin{eqnarray}}
\newcommand{\eea}{\end{eqnarray}}
\newcommand{\bml}{\begin{mathletters}}
\newcommand{\eml}{\end{mathletters}}

\begin{document}

\tighten

\draft



\title{Deformed black strings in 5-dimensional Einstein-Yang-Mills theory}
\renewcommand{\thefootnote}{\fnsymbol{footnote}}
\author{Yves Brihaye\footnote{yves.brihaye@umh.ac.be}}
\affiliation{Facult\'e des Sciences, Universit\'e de Mons-Hainaut, 7000 Mons, Belgium }
\author{Betti Hartmann\footnote{b.hartmann@iu-bremen.de}}
\affiliation{School of Engineering and Sciences, International University Bremen (IUB),
28725 Bremen, Germany}

\date{\today}

\begin{abstract}
We construct the first examples of  
deformed non-abelian black strings in a 5-dimensional Einstein-Yang-Mills
model. Assuming all fields to be independent of the extra coordinate,
we construct deformed black strings, which in the 4-dimensional
picture correspond to  axially symmetric
non-abelian black holes in gravity-dilaton theory.
These solutions thus have {\it deformed} $S^2\times {\mathbb R}$ horizon topology.
We study fundamental properties of the black strings and find
that for all choices of the gravitational coupling
two branches of solutions exist. The limiting behaviour of
the second branch of solutions however depends strongly on the
choice of the gravitational coupling.

\end{abstract}

\maketitle

\section{Introduction}
Higher dimensional black holes have gained a lot of interest
in recent years. This is mainly due to the on-going
study of theories in higher dimensions such as Kaluza-Klein
theories \cite{kaluza,klein}, (super)string theories \cite{pol}
and brane world models \cite{brane}. In higher dimensions,
horizon topologies other than those known in 4 dimensions
are possible. While the first examples of higher-dimensional
black holes, namely the higher dimensional
generalisations of Schwarzschild-, Reissner-Nordstr\"om- \cite{tan}
and Kerr solutions \cite{mp} have horizon topology $S^{d-2}$
in $d$ dimensions, so-called black string solutions \cite{hs}
(in its simplest version
a 4-dimensional Schwarzschild black hole extended into one extra
dimension) with horizon topology $S^2\times {\mathbb R}$ as well as
black ring solutions \cite{er} with horizon topology  $S^2\times S^1$
have been constructed. The black strings (in their simplest version)
are -like
the 5-dimensional Schwarzschild black holes -
static solutions of the vacuum Einstein equations. They have been
mainly investigated with respect to their stability \cite{gl}. Since
the black strings have entropy proportional to $M^{2}/y_0$, where
$M$ denotes the mass of the black string and $y_0$ is the extension in
the extra dimension, hyperspherical black holes have
entropy proportional to $M^{3/2}$. Thus for large enough $y_0$, one would
expect an instability -thereafter called the ``Gregory-Laflamme instability''
- of the black string which was confirmed analytically in \cite{gl}.
For recent reviews on black strings and black holes in space-times
with compact extra dimensions see \cite{kol,ho}.

The higher dimensional Kerr solutions, to which in the
following we refer to as the Myers-Perry solutions, as well as the
Emparan-Reall black ring  solutions are stationary solutions of the vacuum
Einstein equations and thus carry angular momentum.
While the Myers-Perry solutions exist only up to
a maximal value of the angular momentum, black rings are
balanced against gravitational collapse by rotation and thus exist
only above a critical value of the angular momentum.

All existing higher dimensional black holes have been studied intensively
with respect to their uniqueness. While uniqueness theorems
for a variety of static black hole solutions have been well establised \cite{gis,rog,kod},
stationary black holes seem to violate uniqueness.
In $d=5$,  Myers-Perry solutions as well as black ring solutions
exist for the same values of the mass and angular momentum and
are thus {\it not} uniquelly characterised by these latter
parameters. Interestingly, the situation changes if 5-dimensional
supersymmetric black holes and black rings are studied \cite{gut, eemr}. 

While most black hole solutions in higher dimensions
have been constructed in different (dilaton-) gravity theories
without additional matter fields, the first examples
of black holes in a 5-dimensional SO(4)-Einstein-Yang-Mills
model have been constructed \cite{bcht}. These black holes
are hyperspherically symmetric generalisations (horizon topology $S^3$)
of the coloured black hole solutions in 4-dimensional SU(2) Einstein-
Yang-Mills theory \cite{bizon}. These solutions show that  - like
in 4 dimensions -
the uniqueness theorems for higher dimensional static
black holes cannot be extended to models involving non-abelian
gauge fields.

In \cite{hartmann} 5-dimensional black strings
in an SU(2) Einstein-Yang-Mills
model which has been
introduced in \cite{volkov} have been constructed. 
These are 4-dimensional, spherically symmetric 
non-abelian black holes extended trivially into one
extra dimension and have thus horizon topology $S^2 \times {\mathbb R}$.

In this paper, we extend these latter results and discuss deformed black strings.
These black strings are axially symmetric black holes in 4 dimensions
extended trivially into one extra dimension. They thus have {\it deformed}
 $S^2 \times {\mathbb R}$ topology. 
Our solutions are translationally invariant - in contrast to the
recently constructed translationally non-uniform black strings \cite{wiseman}.
To distinguish the case studied here from the (non-) uniform black strings,
we call our solutions with rotational symmetry in $4$ dimensions ``undeformed''
black strings and the solutions with axial symmetry in $4$ dimensions
``deformed'' black strings, respectively.
Note also that the solutions
 studied here are the black hole analogues of the deformed
 vortex-type solutions studied in \cite{bhr}.

 Our paper is organised as follows: In section II, we give the 
 model, the ansatz, the equations of motion and the boundary conditions.
 In Section III, we introduce fundamental properties of the black strings.
 In Section IV, we discuss our numerical results and in Section V, we give our
conclusions.

\section{The Model}
We study the model introduced in \cite{volkov}. This is
an SU(2) Einstein-Yang-Mills model in 5 dimensions, where
all fields are assumed to be independent of the extra dimension.
We take the length of the extra dimension to be equal to unity.

The Einstein-Yang-Mills Lagrangian
in $d=(4+1)$ dimensions then reads:

\begin{equation}
\label{action}
  S = \int \Biggl(
    \frac{1}{16 \pi G_{5}} R   - \frac{1}{4 }F^a_{M N}F^{a M N}
  \Biggr) \sqrt{g^{(5)}} d^{5} x
\end{equation}
with the SU(2) Yang-Mills field strengths
$F^a_{M N} = \partial_M A^a_N -
 \partial_N A^a_M + e\epsilon_{a b c}  A^b_M A^c_N$
, the gauge index
 $a=1,2,3$  and the space-time index
 $M=0,...,4$. $G_{5}$ and $e$ denote
respectively the $5$-dimensional Newton's constant and the coupling
constant of the gauge field theory. $G_{5}$ is related to the Planck mass
$M_{pl}$ by $G_{5}=M_{pl}^{-3}$ and $e^2$ has the dimension of 
$[{\rm length}]$.

Both the metric and matter fields
are assumed to be independent of the extra coordinate $y$. The gauge
fields can then be 
parametrized as follows \cite{volkov}:
\begin{equation}
\label{gauge}
A_M^{a}dx^M=A_{\mu}^a dx^{\mu}+ A_y^a dy   \  , \
\end{equation}
while we give the generalized Ansatz for the metric below.

\subsection{The Ansatz}
Our aim is to construct non-abelian black strings, which are
axially symmetric in 4 dimensions and extended
trivially into one extra dimension. We thus have
three Killing vectors $\frac{\partial}{\partial y}$,
 $\frac{\partial}{\partial \varphi}$, $\frac{\partial}{\partial t}$
associated with the black string solutions. Due to the fact
that we will choose the components of $A_y$ and $A_{\varphi}$ to
point in the same direction of the internal space, off-diagonal
components of the energy-momentum tensor appear. We thus choose the
following ansatz for the metric tensor \cite{bhr}:
\begin{equation}
g^{(5)}_{MN}dx^M dx^N = 
e^{-\xi}\left[-fdt^2+\frac{m}{f}dr^2+\frac{mr^2}{f} d\theta^2+
\frac{l}{f} r^2 \sin^2\theta\left(d\varphi+J dy\right)^2  \right]
+e^{2\xi} dy^2 
\label{metric}
\ , \end{equation}
where $f=f(r,\theta)$, $m=m(r,\theta)$, $l=l(r,\theta)$, $J=J(r,\theta)$ and
$\xi=\xi(r,\theta)$ are functions of $r$ and $\theta$ only.
 We have
parametrized the metric such that the determinant of the metric $g$ becomes
independent of $J$:
\begin{equation}
\sqrt{-g}=e^{-\xi}\frac{mr^2}{f} \sqrt{l} \sin\theta
\end{equation}
Note that our parametrisation here differs from that used
in \cite{bhr}, however can be obtained by a simple transformation
of the fields.

For the gauge fields, the Ansatz reads \cite{bhr}:
\begin{eqnarray}
\label{ansatz1}
A_{\mu}dx^{\mu}&=&
\frac{1}{2er}\left[\tau_{\varphi}^{n}
\left(H_1dr+(1-H_2)rd\theta\right)-n\left
(\tau_r^n H_3 +\tau_{\theta}^n (1-H_4)\right) r\sin\theta d\varphi \right.
\nonumber \\
&+& \left. \left(H_5\tau_r^n+H_6\tau_{\theta}^n\right)rdy \right] 
\ , \end{eqnarray}
where $H_i=H_i(r,\theta)$, $i=1,..,6$ and $\tau_r^n$, $\tau_{\theta}^n$ and
$\tau_{\varphi}^n$ denote the scalar product of the vector of Pauli
matrices $\vec{\tau}=(\tau_1,\tau_2,\tau_3)$ with the unit vectors
$\vec{e}_r^n=(\sin\theta\cos n\varphi, \sin\theta \sin n\varphi, \cos \theta)$,
$\vec{e}_{\theta}^n=(\cos\theta\cos n\varphi, \cos\theta \sin n\varphi, -\sin \theta)$,
$\vec{e}_{\varphi}^n=(-\sin n\varphi, \cos n\varphi, 0)$. $n$
corresponds to the winding number of the configuration.

We fix the residual gauge invariance by imposing the gauge condition
$r\partial_r H_1-\partial_{\theta} H_2=0$ \cite{hkk,bhr}.
\subsection{Equations of motion}

The matter Lagrangian in terms of the field strength tensor reads:
\begin{eqnarray}
{\cal L}_M&=&-\frac{1}{4} {\rm trace}\left( F_{MN} F^{MN}\right)
=-\frac{1}{2} \ {\rm trace} \ 
\left[F_{r\theta}^2 e^{2\xi}\frac{f^2}{m^2 r^2}+
F_{r\varphi}^2 \left(e^{2\xi}\frac{f^2}{ml \sin^2\theta r^2} 
+e^{-\xi} \frac{J^2 f}{m}\right) \right. \nonumber \\
&+& \left.
F_{ry}^2 e^{-\xi} \frac{f}{m}
+  F_{\theta\varphi}^2 \left(e^{2\xi}\frac{f^2}{ml \sin^2\theta r^4} 
+e^{-\xi} \frac{J^2 f}{m r^2}\right)
+F_{\theta y}^2  e^{-\xi} \frac{f}{m r^2} \right. \nonumber \\
&+& \left. F_{\varphi y}^2 \left(e^{-\xi}\frac{f}{\sin^2\theta l r^2}
+ J^2 e^{-4\xi}\right) 
- 2 F_{r\varphi} F_{ry} \frac{e^{-\xi} J f}{m}
- 2 F_{\theta\varphi} F_{\theta y}\frac{e^{-\xi} J f}{m r^2} -
F_{\varphi y}^2 e^{-4\xi} J^2 \right]
\end{eqnarray}
where the non-vanishing parts of the field strength tensor are given by:
\begin{eqnarray}
F_{r\theta}&=&-\frac{1}{r}\left(H_{1,\theta}+
rH_{2,r}\right)\frac{\tau_{\varphi}^n}{2}  \ , \nonumber \\
F_{r\varphi}&=&-n\frac{\sin\theta}{r}\left
(rH_{3,r}-H_1H_4\right)\frac{\tau_{r}^n}{2}+
n\frac{\sin\theta}{r}\left(rH_{4,r}+H_1H_3+\cot\theta H_1\right)
\frac{\tau_{\theta}^n}{2} \ ,   \nonumber \\
F_{\theta\varphi}&=&-n\sin\theta\left
(H_{3,\theta}-1+H_2H_4+\cot\theta H_3\right)\frac{\tau_{r}^n}{2}+
n\sin\theta\left(H_{4,\theta}-H_2H_3-\cot\theta (H_2-H_4)\right)
\frac{\tau_{\theta}^n}{2} \ ,   \nonumber \\
F_{ry}&=&\left
(H_{5,r}+\frac{H_1H_6}{r}\right)\frac{\tau_{r}^n}{2}+
\left( H_{6,r}-\frac{H_1 H_5}{r}\right)
\frac{\tau_{\theta}^n}{2} \ ,    \nonumber \\
F_{\theta y}&=&\left
(H_{5,\theta}-H_2 H_6\right)\frac{\tau_{r}^n}{2}+
\left( H_{6,\theta}+ H_2H_5\right)
\frac{\tau_{\theta}^n}{2} \ ,   \nonumber \\
F_{\varphi y}&=&n\left
(H_{3}H_6 \sin\theta+H_6\cos\theta +H_4 H_5 \sin\theta\right)
\frac{\tau_{\varphi}^n}{2} \ .   \nonumber \\
\end{eqnarray}
The energy-momentum tensor\

\begin{equation}
T_{MN}=2 \ {\rm trace} \ \left(g^{AB} F_{MA} F_{NB} -\frac{1}{4} g_{MN} F_{AB} F^{AB}\right)
\end{equation}
has non-vanishing components $T_{MM}$, $M=0,..,4$ and $T_{r\theta}$, $T_{\varphi y}$. 

The Euler-Lagrange equations $\nabla_M F^{MN}+ i[A_M, F^{MN}]=0$
are obtained by varying the
Lagrangian with respect to the matter fields $H_i(r,\theta)$,
while the
Einstein equations read: $G_{MN}=8\pi G_5 T_{MN}$.
We thus obtain a system of 11 coupled partial differential equations
to be solved subject to appropriate boundary conditions.

Note that in the 4-dimensional picture
our system corresponds to an SU(2) Einstein-Yang-Mills-Higgs-dilaton
system with an additional U(1) potential given in terms of
$J(r,\theta)$ \cite{bhr}.
The $A_y$-component of the gauge field then plays the role of a Higgs
field, while $\xi$ can be interpreted as a dilaton.

\subsection{Boundary conditions}

Due to the requirement of asymptotic flatness, we have for the metric functions
at infinity:
\begin{equation}
f(r=\infty)=1 \ , \ m(r=\infty)=1 \ ,
\ l(r=\infty)=1 \ , \ \xi(r=\infty)=0 \ , \ J(r=\infty)=0 \ ,
\end{equation}
while for the gauge field functions, we have:
\begin{equation}
H_i(r=\infty)=0 \ , \ i=1,2,3,4,6 \ , \ H_5(r=\infty)=1 \ .
\end{equation}
At the regular horizon, the boundary conditions read:
\begin{equation}
f(r=r_h)=0 \ , \ m(r=r_h)=0 \ , \ l(r=r_h)=0 \ , \ \ \left.
\partial_r\xi\right|_{r=r_h}
=0 \ , \ \partial_r J\vert_{r=r_h}=0
\end{equation}
for the metric functions and
\begin{equation}
H_1(r=r_h)=0 \ , \ \partial_r H_i|_{r=r_h}=0 \ , \ i=2,3,4,5,6 \
\end{equation}
for the gauge field.
Finally, the boundary conditions on the $\rho$- and $z$-axis read (due to
symmetry requirements):
\begin{equation}
\partial_{\theta} f|_{\theta=\theta_0}= \partial_{\theta} m|_{\theta=\theta_0}=
\partial_{\theta} l|_{\theta=\theta_0}= \partial_{\theta} \xi
|_{\theta=\theta_0}=\partial_{\theta} J|_{\theta=\theta_0}=  0
\ , \ \theta_0=0, \frac{\pi}{2}
\end{equation}
for the metric fields and
\begin{equation}
H_1(\theta=\theta_0)=H_3(\theta=\theta_0)=H_6(\theta=\theta_0)=
\partial_{\theta} H_2|_{\theta=\theta_0}=\partial_{\theta} H_4|_{\theta=\theta_0}=
\partial_{\theta} H_5|_{\theta=\theta_0}=0 \ , \ \theta_0=0, \frac{\pi}{2}
\end{equation}
for the gauge fields.

\section{Fundamental properties of the black strings}
With 
the introduction of the new variable  $x= r e$ (with $x_h\equiv
r_h e$)
the equations of motion depend only on the coupling constant
\begin{equation}
\alpha^2=4\pi G_{5}   \ .
\end{equation}

The entropy $S$ of the black strings  
is given by:
\begin{equation}
S=\frac{A}{4}=\frac{1}{4}\int\limits_0^{y_0} \int\limits_0^{\pi} 
\int\limits_0^{2\pi} \sqrt{g_{yy}} \ \sqrt{g_{\theta\theta}} \ 
\sqrt{g_{\varphi\varphi}} \ d\varphi \ d\theta \ dy
=\left. y_0 \frac{\pi}{2} \left( \int\limits_{0}^{\pi} d\theta
\sin\theta \sqrt{l \ m}\frac{x^2}{f}\right)\right|_{x=x_h} \ ,
\end{equation}
where $y_0$ is the length of the extra dimension, which
we set to one, $y_0=1$. 

The parameter entering our boundary conditions
is the horizon parameter $x_h$.
However, the interpretation of this parameter is not as straightforward
as in the case of Schwarzschild-like coordinates. We thus use the
area parameter $x_{\Delta}$ \cite{hkk} with
\begin{equation}
x_{\Delta}=\sqrt{\frac{A}{4\pi y_0}}
\end{equation}
to characterise the solutions.

The mass $M$  (per unit length of the extra dimension)
is given by \cite{hkk}:
\begin{equation}
M=\frac{1}{2\alpha^2 }\lim_{x\rightarrow \infty} x^2 \partial_x f \ .
\end{equation}

Furthermore, we study the ratio of the circumference along the equator $L_e$ and that
along the poles $L_p$ to have a measure for the deformation of the 
horizon of the black strings \cite{hkk}:
\begin{equation}
L_e= \left. \left(\int\limits_0^{2\pi} d \varphi \sqrt{\frac{l}{f}} \
\sin\theta \ e^{-\xi/2} \ x\right)\right|_{x=x_h,\theta=\pi/2} 
\ \ \ \ , \ \ \ L_p=2 \left. \left(
\int\limits_0^{\pi} d\theta \  \sqrt{\frac{m}{f}} \  e^{-\xi/2} \ x  
\right)\right|_{x=x_h, \varphi=0}
\end{equation}
A further important property of black holes and black strings is the
temperature of the solutions, which here is given by:
\begin{equation}
\label{temperature}
T=\frac{f_2(\theta)}{2\pi x_h\sqrt{m_2(\theta)}} \ ,
\end{equation}
where we have used the expansion of the metric functions at $x_h$ \cite{hkk}:
\begin{eqnarray}
f(x,\theta)&=&f_2(\theta)\left(\frac{x-x_h}{x_h}\right)^2 + O\left(\frac{x-x_h}{x_h}\right)^3 \ , \nonumber \\
m(x,\theta)&=&m_2(\theta)\left(\frac{x-x_h}{x_h}\right)^2 + O\left(\frac{x-x_h}{x_h}\right)^3 \ .
\end{eqnarray}

The zeroth law of black hole mechanics states that the temperature is constant at the 
horizon, i.e. $\partial_{\theta} T=0$, which requires $2 m_2 \
\partial_{\theta} f_2 - f_2 \ 
\partial_{\theta} m_2 =0$ (this follows directly from (\ref{temperature})).

In the isolated horizon framework for 4-dimensional black hole solutions, it has been
stated \cite{acs} and in fact confirmed for 4-dimensional non-abelian
black holes in SU(2) Einstein-Yang-Mills-Higgs theory \cite{hkk} that a non-abelian
black hole is a bound system of a Schwarzschild black hole and
the corresponding non-abelian regular solution. To test whether this also holds true here, we have studied the
binding energy  $E_b$ (per unit length of the extra dimension) of the black string solutions:
\begin{equation}
E_{b}=M-M_{reg}-M_{s}  \ \ \ ,  \ \ M_s=\frac{x_{\Delta}}{2\alpha^2}
\end{equation}
where $M_s$ is the mass of the Schwarzschild black string (per unit length
of the extra dimension) and $M_{reg}$ is the mass of the corresponding
non-abelian regular solution which has been studied in \cite{bhr}.
Note that for $M_{reg}$, we have used the mass of the fundamental solution,
i.e. the solution on the 1. branch of regular solutions, which
we believe to be stable.

In the study of (non-) uniform black strings and black holes in space-times
with extra compact dimensions, a further quantity, namely the
tension along the extra dimensions has been studied \cite{kol,ho}.
The phase diagram in the mass-tension plane gives good indication
about the properties of the solutions. The detailed study of these 
diagrams has been done in a follow-up publication by the present authors \cite{bhnew}.

\section{Numerical results}
We have solved numerically the system of partial differential equations
subject to the above given boundary conditions
for several values of the coupling constant $\alpha$ and of the
horizon parameter $x_h$, respectively.
Here, we report on our analysis of the cases $n=1$, $n=2$ and $\alpha=0.5$.
More details for other
parameter values will be presented elsewhere \cite{bhnew}.

Before we discuss the numerical results, let us recall the results
for the regular case. It turns out that these are crucial for the understanding
of the qualitative features of the black hole solutions.
The regular case for $n=1$ was studied in detail in \cite{volkov, bh2}.
It has been found that several branches of solutions (which in
the following we refer to as $\alpha$-branches)
for varying $\alpha$ exist. $\alpha$-branch
refers here to a curve giving one of the quantities of the
solution (e.g. the energy) as function of $\alpha$. Typically,
several distinct curves (``branches'') appear in an energy-$\alpha$-plot,
such that for a fixed value of $\alpha$ different solutions (with different
energies) exist.
In \cite{bh2} four branches have been
constructed such that for $\alpha\in [0:0.312[ $,
$\alpha\in\ ]0.419:1.268]$, $\alpha\in[0.312:0.395[$ and
$\alpha\in [0.395:0.419]$ one, two, three and four solutions, respectively,
exist. It is likely that for $\alpha\in [0.395:0.419]$ further branches
exist, however, these have not been constructed so far.

The regular $n=2$ case was studied in \cite{bhr} and only one branch of
solutions in $\alpha$ has been constructed. Corresponding to the
$n=1$ case we believe, though, that further branches also
exist for $n=2$, the numerical construction of which however
seems very involved.

\subsection{Undeformed black strings for $n=1$}
In \cite{hartmann}, the existence of several branches for a fixed
area parameter $x_{\Delta}$ and varying $\alpha$
has been demonstrated for the $n=1$ black strings. The behaviour
of the solutions for fixed $x_{\Delta}$ and varying $\alpha$
is thus similar to that observed for regular solutions \cite{volkov}.
Here, we observe a new phenomenon for $\alpha$ fixed and varying
$x_{\Delta}$. Since this case has not been studied in \cite{hartmann},
we have reconsidered the case $n=1$ here. Note that for $n=1$, the black strings
are 4-dimensional spherically symmetric black holes extended trivially
into the extra dimension. The solutions thus depend only on the
radial coordinate $x$. For the functions we have: $H_2(x)=H_4(x)$,
$H_1(x)=H_3(x)=H_6(x)=J(x)=0$.

\begin{figure}[!htb]
\centering
\leavevmode\epsfxsize=15.0cm
\epsfbox{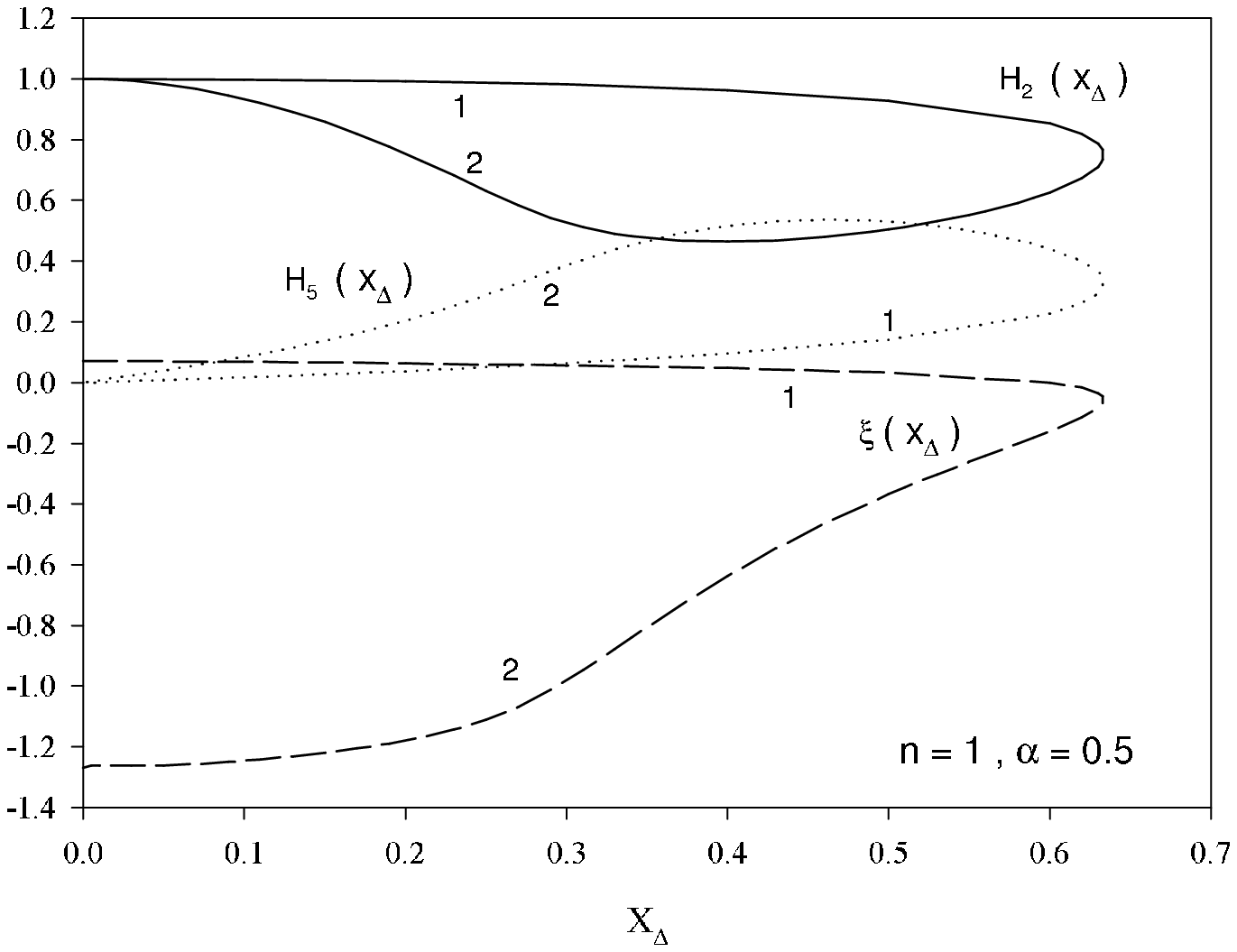}\\
\caption{\label{fig1}
The value of the gauge field functions $H_2=H_4$, $H_5$ and of the
metric function $\xi$ at $x_{\Delta}$, $H_2(x_{\Delta})$, $H_4(x_{\Delta})$,
$\xi(x_{\Delta})$, on the
first (1) and second branch (2) of solutions are shown as function of the 
area parameter $x_{\Delta}$
for the $n=1$ black strings with $\alpha=0.5$.}
\end{figure}

\begin{figure}[!htb]
\centering
\leavevmode\epsfxsize=15.0cm
\epsfbox{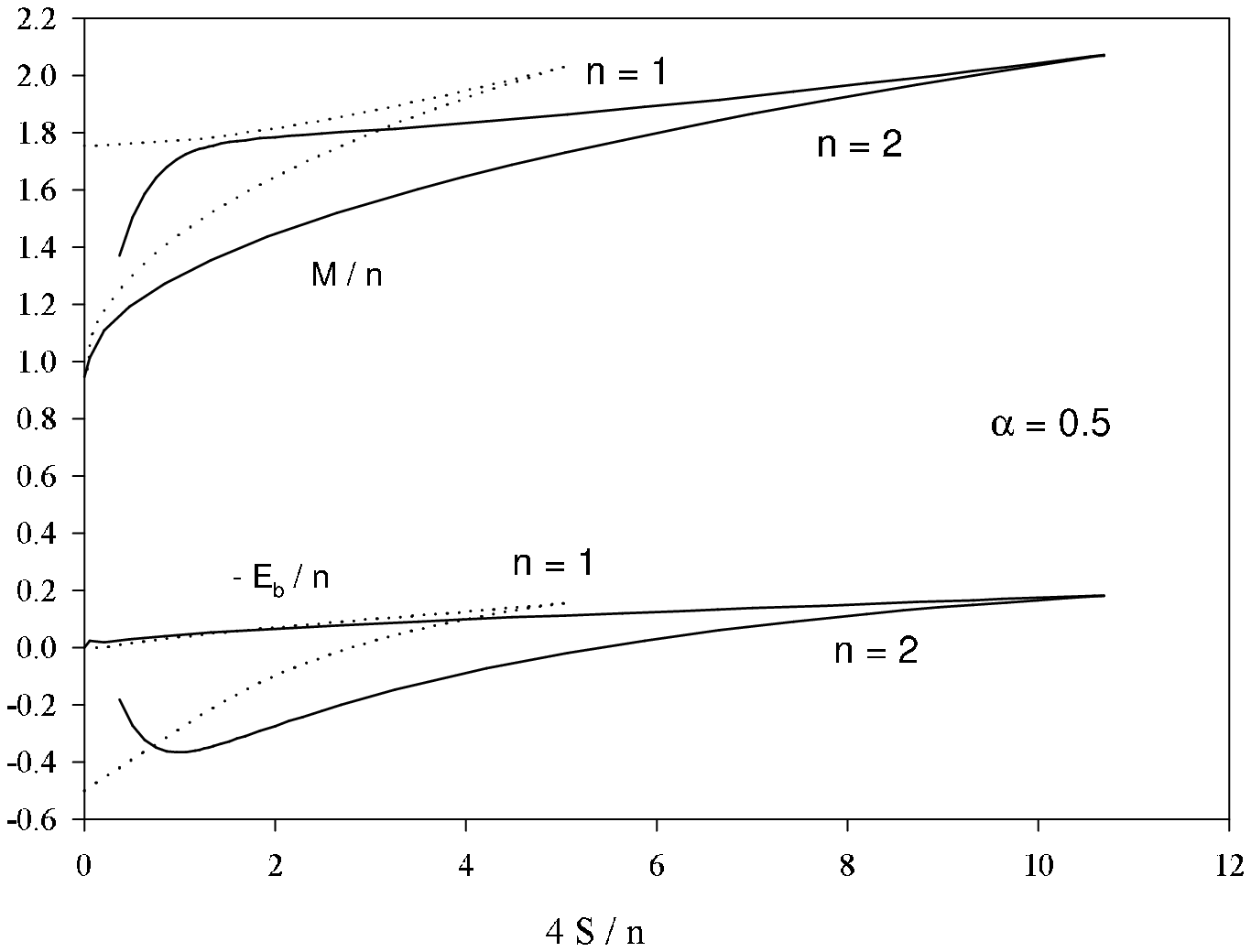}\\
\caption{\label{fig2}
The mass per winding number $M/n$ of the solutions  is shown as function
of (4 times) the entropy per winding number $4S/n$ for the $n=1$ and
$n=2$ black string solutions with $\alpha=0.5$. Also shown
is (the negative of) the binding energy per winding number $-E_b/n$
as function of $4S/n$.
For this latter values, note the inversion of the branches with respect to the plot of the mass.}
\end{figure}

Our numerical results for $\alpha=0.5$ are shown in Fig.\ref{fig1} and
Fig.\ref{fig2}. In Fig.\ref{fig1}, we give the
values of the gauge field functions $H_2(x)=H_4(x)$, $H_5(x)$
and of the metric function $\xi(x)$ at $x_{\Delta}$,
$H_2(x_{\Delta})$, $H_5(x_{\Delta})$, $\xi(x_{\Delta})$,
as functions of the area parameter $x_{\Delta}$. Clearly, two branches
of solutions exist. The first branch (denoted by ``1'')
exists for $x_{\Delta}\in [0:x_{\Delta}^{(max)}]$ with
$x_{\Delta}^{(max)}\approx 0.633$. The limit $x_{\Delta}\rightarrow 0$ on this
branch of solutions corresponds to the fundamental regular solution,
i.e. the solution on the first $\alpha$-branch \cite{volkov}.  
Clearly $H_2(x_{\Delta}\rightarrow 0)\rightarrow
1$ and $H_5(x_{\Delta}\rightarrow 0)\rightarrow
0$, which corresponds to the boundary conditions for
the globally regular solutions at $x=0$. At the same time
$\xi(x_{\Delta}\rightarrow 0)\rightarrow
0.07216$, which is the numerically determined value for
the fundamental regular solution \cite{volkov}. Similarly
the mass on the first (lower) branch tends to the mass
of the fundamental regular solution for $x_{\Delta}\rightarrow 0$
(see Fig. \ref{fig2}).

The second branch of solutions (denoted by ``2'' in Fig.\ref{fig1})
similarly exists for $x_{\Delta}\in [0:x_{\Delta}^{(max)}]$.
Clearly, the solutions on this second branch are distinct from those
on the first branch having higher mass (see Fig.\ref{fig2})
and different values of $H_2(x_{\Delta})$, $H_5(x_{\Delta})$ 
and $\xi(x_{\Delta})$ (see Fig.\ref{fig1}).
In the limit $x_{\Delta}\rightarrow 0$, we find again
that $H_2(x_{\Delta}\rightarrow 0)\rightarrow
1$ and $H_5(x_{\Delta}\rightarrow 0)\rightarrow
0$, while $\xi(x_{\Delta}\rightarrow 0)\rightarrow -1.262$ tends to the
value of the corresponding solution of the second $\alpha$-branch of
regular solutions. Strong evidence for this is
also given by inspection of the mass curve in Fig.\ref{fig2}, where
in the limit $x_{\Delta}\rightarrow 0$, the mass tends
to that of the regular solution of the second $\alpha$-branch.
 
The binding energy $E_b$ (see Fig.\ref{fig2}) is negative on the first branch, indicating
that indeed non-abelian black strings are bound systems of
a Schwarzschild black string and the corresponding regular
non-abelian vortex solution. On the second branch, the binding energy becomes
positive, indicating an instability of the solutions.
Note that for both branches, we have used the
mass of the fundamental regular solution, which we believe to be
the stable regular solution. If we had used the
mass of the regular solution on the second $\alpha$-branch to obtain
the binding energy for our black string solutions on the second
branch, the binding energy would have been negative.

Results for different values of $\alpha$ will be given elsewhere \cite{bhnew}.
However, we believe that for all values of $\alpha$ two branches of black string
solutions will exist. The critical behaviour, however, will strongly  
depend on the value of $\alpha$. For $\alpha=0.5 \in ]0.419:1.268]$
two regular solutions exist and the second branch terminates in the
corresponding regular solution of the second $\alpha$-branch.
We believe that if $\alpha \in [\alpha_1^{(n)}:\alpha_2^{(n)}]$, where
$n$ indicates the number of regular solutions available for this
range of $\alpha$, i.e. the number of branches, 
the black string solutions on the second branch
will tend to the $n$th regular solution, i.e. to the solution
on the $n$th branch, for $x_{\Delta}\rightarrow 0$. 
E.g. the black strings on the second branch
of solutions for $\alpha=0.35\in [0.312:0.395[$ would tend to the
third solution, i.e. the solution on the third $\alpha$-branch of regular
solutions.
For $\alpha\in\ ]0:0.312]$ for which only the fundamental regular solution
exists, we expect that the second branch terminates at some {\it finite}
$x_{\Delta} > 0$. At this point, the solution
bifurcates with the branch of Einstein-Maxwell-dilaton solution 
with $H_2(x)=H_4(x)\equiv 0$, $H_5(x)\equiv 1$ and $\xi(x)$ given
by the corresponding function of the Einstein-Maxwell-dilaton solution.
      
\subsection{Deformed black strings for $n=2$}
\begin{figure}[!htb]
\centering
\leavevmode\epsfxsize=15.0cm
\epsfbox{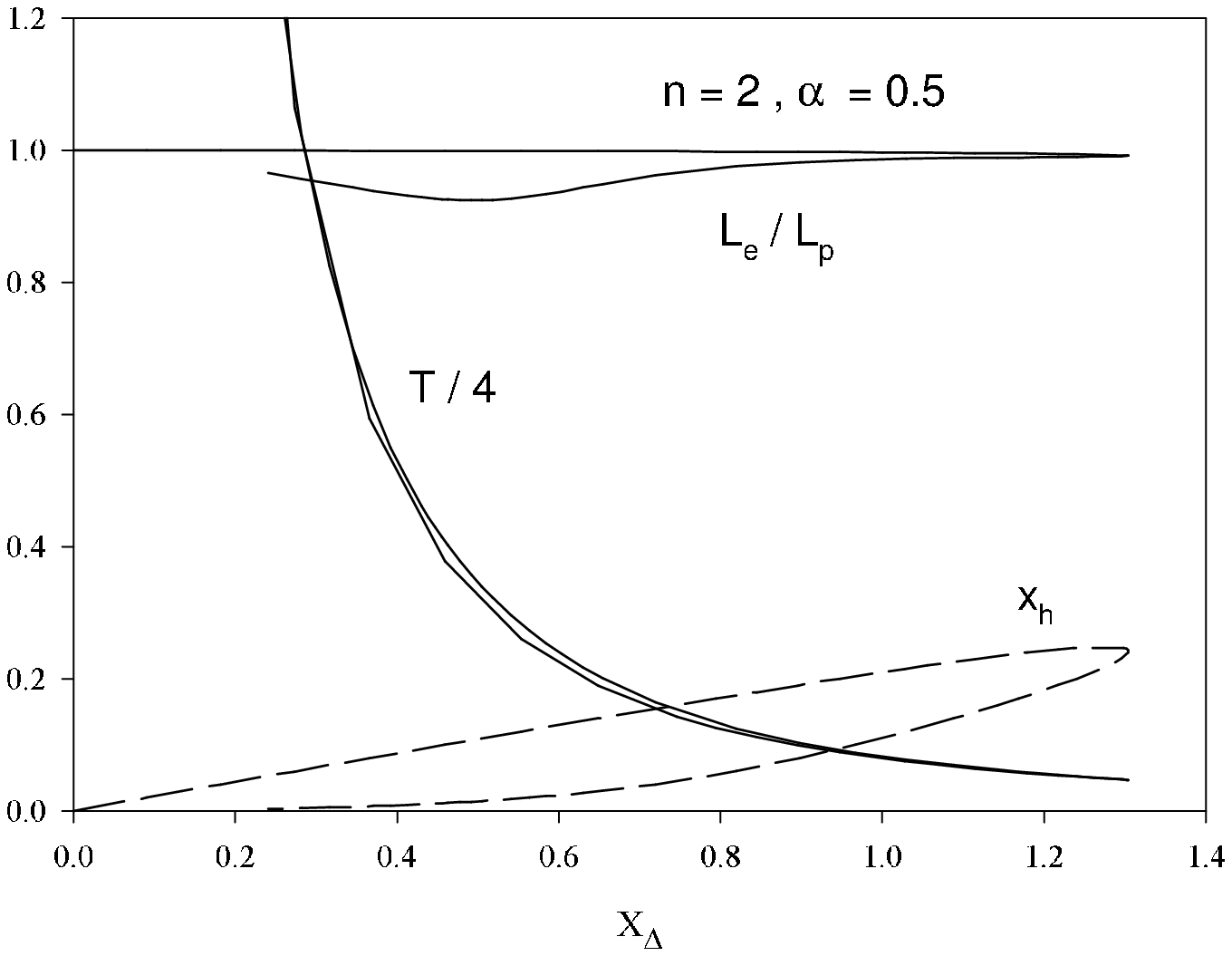}\\
\caption{\label{fig3}
The ratio of the circumference
almong the equator and the 
circumference along the poles $L_e/L_p$, the temperature
$T$
as well as the horizon parameter $x_h$ are shown as functions
of the area parameter $x_{\Delta}$ for the deformed
black string solutions with $n=2$ and $\alpha=0.5$.}
\end{figure}

\begin{figure}[!htb]
\centering
\leavevmode\epsfxsize=15.0cm
\epsfbox{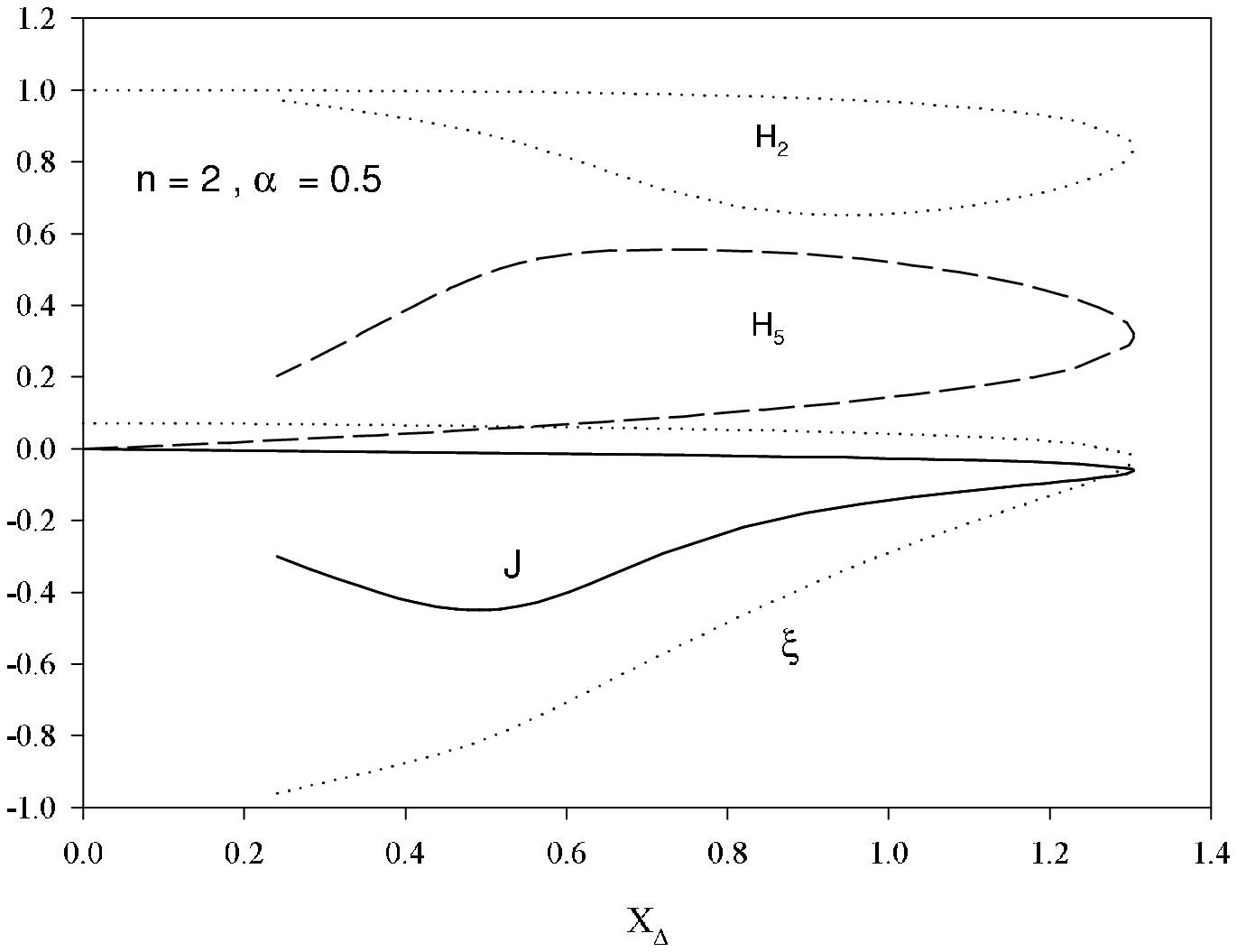}\\
\caption{\label{fig4}
The values of the metric functions $\xi$ and $J$ and of the
gauge field functions $H_2$ and $H_5$ at $x_{\Delta}$,
$\xi(x_{\Delta},\theta=0)$, $J(x_{\Delta},\theta=0)$,
$H_2(x_{\Delta},\theta=0)$,$H_5(x_{\Delta},\theta=0)$,
are shown as functions of the area parameter $x_{\Delta}$
for the black string solutions with $n=2$ and $\alpha=0.5$.}
\end{figure}

Our results for the deformed black string solutions ($n=2$) 
are given in Fig.\ref{fig2}, Fig.\ref{fig3} and Fig.\ref{fig4}.

The first branch exists for $x_{\Delta} \in [0:x_{\Delta,max}]$
with $x_{\Delta,max}\approx 1.31$. In the limit 
$x_{\Delta}\rightarrow 0$, the solution approaches the
corresponding regular solution which has been
constructed in \cite{bhr}. When increasing $x_{\Delta}$, our results
show that both the mass $M$ (see Fig.\ref{fig2})
and the corresponding value of the horizon parameter $x_h$ increase,
while the temperature $T$ decreases (see Fig.\ref{fig3}). We have confirmed numerically
that the temperature on the horizon is constant and our solutions
thus fulfill the zeroth law of black hole mechanics.
The deformation parameter $L_e/L_p$ decreases, but stays very close to
one indicating that the horizon is only deformed slightly. On this branch,
the absolute value of the new function $\vert J(x,\theta)\vert$
stays small. In Fig.\ref{fig4}
we show the values of $J$ at the horizon, $J(x_{\Delta}, \theta=0)$ 
together with the values of $\xi(x_{\Delta}, \theta=0)$.
These latter values are positive on the first branch of solutions.
The values $H_2(x_{\Delta},\theta)$ and
$H_4(x_{\Delta},\theta)$ decrease from one, while
$H_5(x_{\Delta},\theta)$ increases from zero for
increasing  $x_{\Delta}$. We demonstrate the $x_{\Delta}$-dependence
for
$H_2(x_{\Delta},\theta=0)$ and $H_5(x_{\Delta},\theta=0)$ in Fig.\ref{fig4}.

Like in the $n=1$ case the limit $x_{\Delta}\rightarrow 0$ corresponds
to the fundamental regular deformed vortex solution \cite{bhr}.

We managed to construct the second branch of solutions
for $x_{\Delta} 
\in [x_{\Delta,end}:x_{\Delta,max}]$ with
$x_{\Delta,end}\approx 0.2$. At
$x_{\Delta}=x_{\Delta,max}$ the branches merge into a single solution.
The mass of the solutions on the
second branch is higher than that of the corresponding solution
on the first branch for the same value of $x_{\Delta}$ and thus
same entropy (see Fig.\ref{fig2}).

The other features of the second branch are that, when $x_{\Delta}$
decreases from $x_{\Delta,max}$, the parameter $x_h$
and the mass decrease, while the temperature $T$ increases
and stays higher than the temperature of the corresponding solution
on the first branch. As compared to the solutions
on the first branch, the solutions on the second branch
have a much stronger deformed horizon.
This can be noted by observing the curve $L_e/L_p$ in Fig.\ref{fig3}
and the data plotted in Fig.\ref{fig4}. We also notice
that the value of the
metric function $\xi$ at the horizon becomes negative
on the second branch and that the value of $J$ at the horizon
now deviates significantly from zero.

Finally, let us mention that the values of  $H_2(x_{\Delta},\theta)$ and
$H_4(x_{\Delta},\theta)$ start to increase again, 
while $H_5(x_{\Delta},\theta)$ decreases for
decreasing $x_{\Delta}$. The detailed curves for $H_2(x_{\Delta},\theta=0)$
and $H_5(x_{\Delta},\theta=0)$ are shown in Fig.\ref{fig4}.

We strongly believe that this second branch extends all the way back 
to $x_{\Delta}=0$
similar to the $n=1$ case. Then $H_2(x_{\Delta}\rightarrow 0,\theta)\rightarrow
1$, $H_5(x_{\Delta}\rightarrow 0,
\theta)\rightarrow 0$, $\xi(x_{\Delta}\rightarrow 0,\theta)\rightarrow
\xi_0 < 0$, $J(x_{\Delta}\rightarrow 0,\theta)\rightarrow
J_0 < 0$, where $\xi_0$ and $J_0$ have not been determined so far.
Correspondingly, $x_h\rightarrow 0$ in this limit, $L_e/L_p \rightarrow 1$,
(like for the first branch) and $T\rightarrow \infty$, since $T$ is not defined
for regular solutions.

In Fig.\ref{fig2}, we also show (the negative of) the binding energy per winding
number  $-E_b/n$ for $\alpha=0.5$. 
Clearly, the binding energy on the first branch of solutions
(note the inversion of the branches with respect to the plot of
the mass) is
negative and the non-abelian black strings on the first branch 
are thus {\it bound systems} of 
the corresponding regular non-abelian vortex solutions of \cite{bhr} 
and the Schwarzschild black string. On the second branch, the situation
changes. The binding energy becomes positive.
This signals that the non-abelian black strings are unstable
to decay into the non-abelian, globally regular vortex solutions 
and a Schwarzschild black string on this second branch of solutions.
 
Comparing the $n=1$ and $n=2$ solutions for the same value of $\alpha$,
we find that the extension of the branches in $x_{\Delta}$ is bigger for 
$n=2$ as compared to  $n=1$. Furthermore, when comparing the
respective first and second branches for $n=1$ and $n=2$, the mass
(per winding number) of the $n=2$ solution is always lower than that of the $n=1$ 
solution.

Results for different values of $\alpha$ will be given elsewhere.
However, we believe that the scenario is similar to the $n=1$ case,
such that the second branch terminates in the corresponding regular
solution of the $n$th $\alpha$-branch for $x_{\Delta}\rightarrow 0$ if more than one $\alpha$-branch exist,
or in an Einstein-Maxwell-dilaton solution for finite $x_{\Delta}$ with $H_1(x,\theta)=H_2(x,\theta)=H_3(x,\theta)=
H_4(x,\theta)=H_6(x,\theta)=0$, $H_5(x,\theta)\equiv 1$, $J(x,\theta)=0$
and $\xi(x,\theta)=\xi(x)$ given by the corresponding function
of the Einstein-Maxwell-dilaton solution if only one $\alpha$-branch
exists for the regular solutions. 
\section{Conclusions}
In this paper, we have constructed black strings as
solutions of a 5-dimensional Einstein-Yang-Mills model.
We have presented our results for $n=1$, $n=2$ and fixed $\alpha=0.5$.
We find that two branches of solutions exist in both cases,
which in the limit of $x_{\Delta}\rightarrow 0$ tend to the corresponding
regular solutions of the first, respectively second $\alpha$-branch.
We believe that this is a generic feature of the system.
For other values of $\alpha$, the second branch of black string
solutions will terminate in the $n$th available regular solution.
This means that in specific parameter ranges, the black string
solutions will start to show an oscillatoric behaviour in the
gauge field functions. For those values of $\alpha$, for which
only one globally regular solution exists, we believe that
the second branch will terminate at a finite value of the
area parameter in an Einstein-Maxwell-dilaton solution.
More details will be given elsewhere \cite{bhnew}.

The question whether the non-abelian black strings are unstable
to decay to hyperspherically symmetric non-abelian black holes, i.e.
would have an instability corresponding
to the ``Gregory-Laflamme instability'' for Schwarzschild black strings
is beyond the scope of this paper. A first step to answer this question
would be the numerical construction of hyperspherically symmetric
non-abelian black holes of the SU(2) Einstein-Yang-Mills system in
5 dimensions.

We remark that next to the ``fundamental'' solutions constructed here,
also excited solutions could exist-similar to what is known for
4-dimensional spherically symmetric black hole solutions in SU(2) Einstein-Yang-Mills-Higgs theory
\cite{bfm}. These solutions would have a number $m$, $m\in \mathbb{N}$,
of zeros of the gauge field functions. The construction of these
solutions has not been done so far - neither in the case of
globally regular vortices nor black strings - and is left to a future
publication.\\
\\
\\
{\bf Acknowledgment} We thank the Belgian F.N.R.S. for financial support.

\newpage

\newpage

\newpage

\newpage

\end{document}